\documentclass[12pt,a4paper]{article}
\usepackage{amsmath,amsthm,amsfonts,amssymb,alltt}

\usepackage{epsfig}
\usepackage{amsmath,amsfonts}

\newtheorem{theorem}{Theorem}
\newtheorem{lemma}[theorem]{Lemma}

\newtheorem{Example}{Example}

\theoremstyle{definition}

\theoremstyle{remark}
\newtheorem{remark}[theorem]{Remark}

\numberwithin{equation}{section}

\begin{document}
\sloppy

\title {Composites with invisible inclusions: eigenvalues of $\mathbb R$-linear problem}

\author{V. V. Mityushev}

\begin{abstract}
An new eigenvalue $\mathbb R$-linear problem arisen in the theory of metamaterials is stated and constructively investigated for circular non-overlapping inclusions. An asymptotic formula for eigenvalues is deduced when the radii of inclusions tend to zero. The nodal domains conjecture related to univalent eigenfunctions is posed. Demonstration of the conjecture allows to justify that a set of inclusions can be made neutral by surrounding it with an appropriate coating.   
\end{abstract}

\maketitle
\section{Introduction}
Local fields in fibrous composites are described by solutions of the Riemann-Hilbert and the $\mathbb R$-linear problems for multiply connected domains \cite{andr2, 3, Mit2012b, Rylko1,  Rylko2, Rylko3}. The physical properties of the components of traditional composites are expressed in terms of the positive constants, c.f., conductivity, permeability, permittivity etc.

Recently, materials having negative physical constants were discovered. It concerns dielectric-magnetic materials displaying a negative index of refraction \cite{Alu, Asatryan, McPh,  McPhedran2015, Poulton}. 
Mathematical modelling of metamaterials and neutral (invisible) inclusions were discussed in \cite{JM, Kerker, MiltonS} and works cited therein. 
In particular, the paper \cite{JM} contains a general observation that any shaped inclusion with a smooth boundary can be made neutral by surrounding it with an appropriate coating. This result is based on the study of the eigenvalues of the $\mathbb R$-linear problem for a doubly connected domain $D$ when the spectral parameter is assigned only to one component of $\partial D$.  Such a problem can be considered as a modification of the result \cite{Sch, Mityushev} devoted to eigenvalues of the $\mathbb R$-linear problem with the same spectral parameter in each component of the boundary. 

The discussed eigenvalue problem differs from the classic problem when the spectral parameter $\lambda$ enters into equation, for instance, $\Delta u + \lambda u =0$  \cite{Chavel}. Our eigenvalue problem is similar to the Steklov problem \cite{ag} when $\Delta u  =0$ in $D$ and $u = \lambda \frac{\partial u}{\partial n}$ on the boundary. Similar mixed boundary-spectral $\mathbb R$-linear problems were studied in \cite{Bogatyrev1}, \cite{Bogatyrev2} by reduction to integral equations and in \cite{Banuelos} by variational methods. 

In the present paper, we state the general eigenvalue $\mathbb R$-linear problem arisen in the theory of metamaterials and investigate it for circular non-overlapping inclusions. An asymptotic formula for eigenvalues is deduced when the radii of inclusions tend to zero. 
The nodal domains conjecture related to univalent eigenfunctions is posed.    

\section{Statement of eigenvalues $\mathbb R$-linear problem}
\label{sec1}
\begin{figure}[ht]
\centering \epsfig{figure=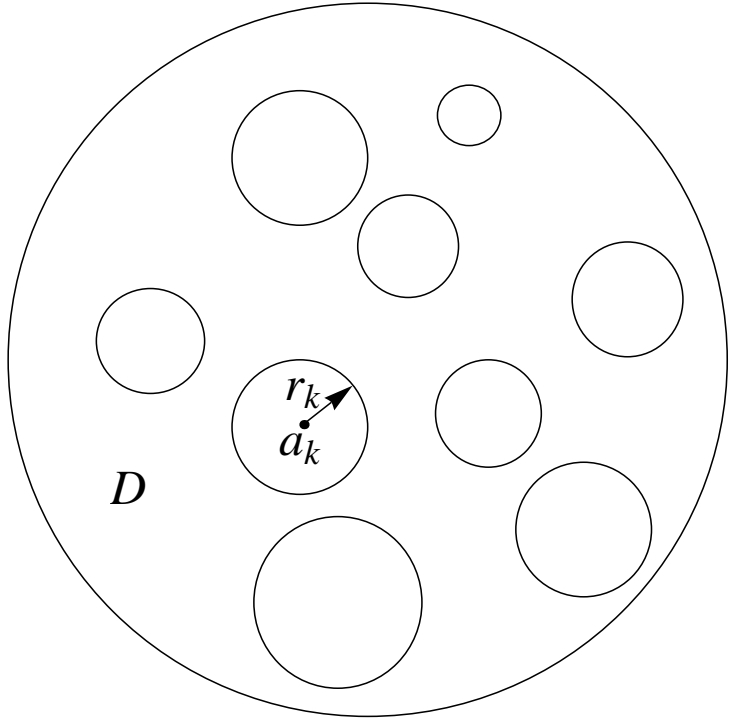, width=.60\linewidth}
\caption{Multiply connected domain $D$ with circular boundaries.}\label{fig1}
\end{figure}

Let $\widehat{\mathbb C}=\mathbb C \cup \{\infty\}$ denote the extended complex plane. Consider $n$ non-overlapping simply connected domains $D_k$ ($k=1,2,\cdots, n$) lying in the unit disk $U$ and the multiply connected domain $D=U\backslash \cup_{k=1}^n (D_k \cup \partial D_k)$ (see Fig.\ref{fig1}). Let $D_0$ denote the exterior of the closed unit disk to the extended complex plane. Let the boundary of each $D_k$ ($k=0,1,\cdots, n$) be a positively oriented smooth simple curve $\Gamma_k$ including the unit circle $\Gamma_0$.

Given H\"{o}lder continuous functions $a_k(t)$, $b_k(t)$ on $\Gamma_k$  satisfying the inequality $|a_k(t)|>|b_k(t)|$ ($k=0,1,2,\cdots, n$). It is assumed that the winding number (index) of each $a_k(t)$ vanishes \cite{Gakhov}. To find functions $\varphi_k(z)$ analytic in $D_k$,  respectively, continuous in the closures of the considered domains and to find a complex constant $\lambda \neq 0$ such that the following $\mathbb R$-linear conditions are fulfilled
 \begin {equation}
\label {R1}
\varphi(t) = a_k(t) \varphi_k(t) +b_k(t) \overline{\varphi_k(t)}, \quad t \in \Gamma_k, \quad k=1,2,\cdots, n,
\end {equation}
 \begin {equation}
\label {R2}
\varphi(t) = \overline{\lambda} \; a_0(t)  \varphi_0(t) +  b_0(t) \overline{\varphi_0(t)}, \quad |t|=1.
\end {equation}
It is assumed that the unknown function $\varphi_0(z)$ is analytic in $|z|>1$ continuous in $|z|\geq 1$ and vanishes at infinity:
 \begin {equation}
\label {R3}
\varphi_0(\infty)=0.
\end {equation}

A non--zero function $\varphi_0(z)$ satisfying \eqref{R1}-\eqref{R3} is called the eigenfunction and the corresponding constant $\lambda$ the eigenvalue of the problem. The function $\varphi_0(z)$ is distinguished from others, since the function $\omega(z)=\overline{\varphi_0 \left(\frac{1}{\overline{z}} \right)}$, $|z| \geq 1$, plays the key role in the theory of metamaterials. The univalent function  $\omega(z)$ determines the shapes of the inclusions $\omega(\Gamma_k)$  ($k=1,2,\ldots, n$) and of the corresponding neutral coating $\omega(\Gamma_0)$.

It follows from Bojarski's theorem \cite{Bojarski, BM} that the eigenvalues of the problem \eqref{R1}-\eqref{R3} satisfy the inequality $|\lambda|\leq 1$.

\section{Functional equations}
\label{sec2}
We consider the problem \eqref{R1}--\eqref{R3} with the constant coefficients $a_k(t) = 1$, $b_k(t) = -\rho_k$, where $|\rho_k|<1$ ($k=1,2,\ldots, n$) and $a_0(t) = 1$, $b_0(t) = -1$. It is also assumed that $\Gamma_k$ are circles $|t-a_k|=r_k$. Then, \eqref{R1}--\eqref{R3} become 
 \begin {equation}
\label{eq:pr1}
\varphi(t) = \varphi_k(t) -\rho_k \overline{\varphi_k(t)}, \quad t \in \Gamma_k, \quad k=1,2,\cdots, n,
\end {equation}
 \begin {equation}
\label{eq:pr2}
\varphi(t) =\overline{\lambda} \varphi_0(t) - \overline{\varphi_0(t)}, \quad |t|=1,
\end {equation}
 \begin {equation}
\label{eq:pr3}
\varphi_0(\infty)=0.
\end {equation}
The problem \eqref{eq:pr1}-\eqref{eq:pr3} can be stated in terms of harmonic functions \cite{Mityushev3, Rylko0}. For instance, the condition \eqref{eq:pr2} for real $\lambda$ up to an additive constant can be written in the form
 \begin {equation}
\label{eq:pr2h}
u =(\lambda-1)u_0, \quad \frac{\partial u}{\partial n} =(\lambda+1)\frac{\partial u_0}{\partial n} , \quad |t|=1,
\end {equation}
where $\frac{\partial}{\partial n}$ denotes the outward normal derivative to the unit circle, $u =$Re$\varphi$ and $u_0 =$Re$\varphi_0$. The $\mathbb R$-linear problem \eqref{eq:pr1}-\eqref{eq:pr3} describes neutral inclusions with the contrast parameters $\rho_k$ ($k=1,2,\ldots, n$) \cite{Mityushev}.

Following \cite{Mityushev3,3} we reduce the problem \eqref{eq:pr1}-\eqref{eq:pr3} to a system of functional equations. Let $$z_{(m)}^{\ast } = \frac{r_m^2}{\overline{z-a_m}}+a_m$$ 
denote the inversion with respect to the circle $|z-a_k|=r_k$.
Introduce the function 
$$
\Phi (z) :=\left\{
\begin{array}{l}
\varphi _{k}(z) + \sum_{m\neq k} \rho_m \overline{\varphi_m \left(
z_{(m)}^{\ast }\right) }+\overline{ \varphi_0 \left(\frac{1}{\overline{z}} \right)}, 
\\
\qquad \qquad \qquad \qquad \qquad \qquad \qquad \quad |z-a_{k}|\leq r_k, \quad k=1,2,\ldots,n,
\\
\varphi(z) + \sum_{m=1}^n \rho_m \overline{\varphi_m \left(
z_{(m)}^{\ast }\right) }+\overline{ \varphi_0 \left(\frac{1}{\overline{z}} \right)},\;z\in D,
\\
\\
\overline{\lambda} \varphi_0(z) + \sum_{m=1}^n \rho_m \overline{\varphi_m \left(
z_{(m)}^{\ast }\right) },\;|z| \geq 1.
\end{array}
\right.
$$
analytic in $D_k$ $(k=0,1, \ldots,n)$   and $D$.
Calculate the jump of $\Phi(z)$ across the circle $|t-a_k|= r_k$
$$
\Delta _{k}:=\Phi ^{+}(t) -\Phi ^{-}(t) ,\;|t-a_{k}|= r_k,
$$
where $\Phi ^{+}\left( t\right) :=\lim_{z\rightarrow t\;z\in
D }\Phi \left( z\right) ,\;\Phi ^{-}\left( t\right)
:=\lim_{z\rightarrow t\;z\in D_{k}}\Phi \left( z\right)
$. Application of \eqref{eq:pr1} gives $\Delta _{k}=0.$ Similar arguments for the jump $\Delta_0$ of $\Phi(z)$ across the unit circle yield $\Delta_0=0$. It follows from the principle of analytic continuation that $\Phi(z)$ is analytic in the extended complex plane. 
Then, Liouville's theorem implies that $\Phi(z)$ is a constant. Calculation of this constant as $\Phi(\infty)$ and using of \eqref{eq:pr3} yields
\begin{equation} 
\Phi(z)= \sum_{m=1}^n \rho_m \overline{\varphi_m \left(a_m\right) }.
\label{eq:Phi}
\end{equation} 
The definition of $\Phi (z)$ in $| z-a_{k}| \leq r_{k}$ and $|z|\geq 1$ leads to the following system of functional equations
\begin{equation}
\begin{array}{lll}
\label{eq:rh81}
\varphi _{k}(z)=- \sum_{m\neq k} \rho_m \left[\overline{\varphi_m \left(
z_{(m)}^{\ast }\right) }-\overline{\varphi_m \left(
a_m\right) }\right]+ \rho_k\overline{\varphi_k(
a_k) } - \overline{ \varphi_0 \left(\frac{1}{\overline{z}} \right)}, 
\\ 
\quad \quad \quad \quad \quad \quad \quad \quad \quad \quad \quad \quad \quad \quad | z-a_{k}| \leq r_{k}, \; k=1,2,\ldots,n,
\\
\\
\overline{\lambda} \varphi_0(z) =- \sum_{m=1}^n \rho_m \left[\overline{\varphi_m \left(
z_{(m)}^{\ast }\right) }-\overline{\varphi_m \left(
a_m\right) }\right],\;|z| \geq 1.
\end{array}
\end{equation}
Exclusion of $\varphi_0(z)$ from \eqref{eq:rh81} yields the system
\begin{equation}
\begin{array}{lll}
\label{eq:rh82}
\varphi _{k}(z)=- \sum_{m\neq k} \rho_m \left[\overline{\varphi_m \left(
z_{(m)}^{\ast }\right) }-\overline{\varphi_m \left(
a_m\right) }\right]+ \rho_k\overline{\varphi_k(
a_k) }+
\\ 
\frac{1}{\lambda} \sum_{m=1}^n  \overline{\rho_m} \left[\varphi_m \left(
a_m+ \frac{r_m^2 z}{1-\overline{a_m}z}\right) - \varphi_m \left(
a_m\right) \right],
\\
\quad \quad \quad \quad \quad \quad \quad \quad \quad \quad \quad \quad \quad \quad | z-a_{k}| \leq r_{k}, \; k=1,2,\ldots,n.
\end{array}
\end{equation}
We will assume that $\varphi _{k}(z)$ are analytic in $| z-a_{k}| <r_{k}$ and continuously differentiable in $| z-a_{k}| \leq r_{k}$ due to the physical treatment of $\varphi _{k}(z)$ as complex potentials. Introduce the space of functions $\mathcal C^1 (\cup_{m=1}^n D_m)$ analytic in the non-connected domain  $\cup_{m=1}^n D_m$ and continuously differentiable in its closure with the norm 
$$\|\phi \|_{\mathcal C^1} = \max_{m=1,2,\ldots,n} \; \max_{| z-a_{m}|= r_{m}} |\varphi_{m}(z)| + \max_{m=1,2,\ldots,n} \; \max_{| z-a_{m}|= r_{m}} |\varphi'_{m}(z)|,$$
where $\phi(z)=\varphi_{m}(z)$ in $| z-a_{m}|\leq r_{m}$.
One can write the system \eqref{eq:rh82} as an equation in the Banach space $\mathcal C^1 (\cup_{m=1}^n D_m)$    
\begin{equation}
\label{eq:rh90}
\phi = A\phi+ \frac{1}{\lambda} B\phi,
\end{equation}
where the operators $A$ and $B$ are introduced in accordance with \eqref{eq:rh82} for shortness. Equation \eqref{eq:rh90} can be considered in the Hilbert space $\mathcal H^2 (\cup_{m=1}^n D_m)$ of functions $\phi(z)=\varphi _{k}(z)$ which belong to the Hardy space in the disks $| z-a_{k}| <r_{k}$ with the norm \cite{Drygas}
$$\|\phi \|_{\mathcal \mathcal H^2} = \left(\sum_{m=1}^n\;  \sup_{R<r_m} \frac{1}{2\pi} \int_0^{2\pi} |\varphi _{m}(a_m + R e^{i\theta})|^2 d\theta \right)^{\frac 12}.$$

It follows from \cite{Mityushev3} that the operators $A$ and $B$ are compact in the considered spaces and the operator $I-A$ is invertible where $I$ denotes the identity operator. Then, equation \eqref{eq:rh90} is equivalent to the eigenvalue problem 
\begin{equation}
\label{eq:rh91}
\lambda \phi = (I-A)^{-1} B\phi,
\end{equation}
where the  operator $(I-A)^{-1} B$ is compact in the space $\mathcal H^2 (\cup_{m=1}^n D_m)$. Therefore, the eigenvalue problem \eqref{eq:rh82} can be written in the form of the eigenvalue problem \eqref{eq:rh91} for a compact operator in the Hilbert space. Let $\phi \in \mathcal H^2 (\cup_{m=1}^n D_m)$ be its eigenfunction. Then, Pumping principle \cite{Mityushev3} implies that $\phi$ actually belongs to $\mathcal C^1 (\cup_{m=1}^n D_m)$ (even to $\mathcal C^\infty$). It is based on the following arguments. For a fixed $k$, every function $\overline{\varphi_m \left(z_{(m)}^{\ast }\right)}$ ($m\neq k$) is analytic in $|z-a_m|>r_m$ and $\overline{ \varphi_0 \left(\frac{1}{\overline{z}} \right)}$ in $|z|<1$. The union of these domains contains the closed disk $|z-a_k|\leq r_k$. Hence, the right part of \eqref{eq:rh82} is analytic in $|z-a_k|\leq r_k$. Therefore, the left part containing the function $\varphi_k(z)$, is also analytic in $|z-a_k|\leq r_k$.

Instead of the functional equations \eqref{eq:rh82} we consider equations in the space $\mathcal C (\cup_{m=1}^n D_m)$ associated with continuous functions obtained by differentiation of \eqref{eq:rh82}
\begin{equation}
\begin{array}{lll}
\label{eq:rh14}
\psi_k(z)= \sum_{m\neq k} \rho_m \frac{r^2_m}{(z-a_m)^2}\; \overline{\psi_m \left(z_{(m)}^{\ast }\right) } +
\\
\\
\frac{1}{\lambda} \sum_{m=1}^n \overline{\rho_m} \frac{r_m^2}{(1-\overline{a_m}z)^2}\; 
\psi_m \left(a_m+ \frac{r^2_m z}{1-\overline{a_m}z}\right) ,
\\
\quad \quad \quad \quad \quad \quad \quad \quad \quad \quad \quad \quad \quad \quad | z-a_{k}| \leq r_{k}, \; k=1,2,\ldots,n,
\end{array}
\end{equation}
where $\psi_k(z) = \varphi'_{k}(z)$. One can see from the second equation \eqref{eq:rh81} that $\varphi_0(z)$ does not depend on $\varphi_m(a_m)$. Therefore, one can first solve the system \eqref{eq:rh14} and determine  
\begin{equation}
\label{eq:phi0}
\varphi'_0(z) = \frac{1}{\overline{\lambda}} \sum_{m=1}^n  \frac{\rho_m r^2_m}{(z-a_m)^2}\;\overline{\psi_m \left(z_{(m)}^{\ast }\right) },\quad|z| \geq 1.
\end{equation}
The function $\varphi_0(z)$ is uniquely found from \eqref{eq:phi0} by integration  
\begin{equation}
\label{eq:phi01}
\varphi_0(z) =-\int_z^{\infty}\varphi'_0(\zeta)d\zeta,\quad|z| \geq 1.
\end{equation}

\begin{remark}
\label{rem1}
The eigenvalues of the Laplace operator form an increasing sequence \cite{Chavel}. In our case, the eigenvalue problem \eqref{eq:rh91} or \eqref{eq:rh14} is addressed to a compact operator. Therefore, the absolute values of eigenvalues decrease to zero \cite{KF}.
\end{remark}

\section{Asymptotic solution of functional equations}
In the present section, we find asymptotic solutions of the systems \eqref{eq:rh81} and \eqref{eq:rh82} when $r=\max_{k =1,2, \ldots,n} r_k$ tends to zero. The parameters $\nu_k = \frac{r_k^2}{r^2}$ are considered as values for which $0<\nu_k \leq 1$ including the limit case, as $r \to 0$.  

\begin{lemma}
\label{lemma1}
The eigenvalues $\lambda = \lambda(r)$ satisfy the asymptotic relation
\begin{equation}
\label{eq:lam}
\lambda(r)= r^2 \lambda_0(r), \quad \mbox{as} \;r \to 0,
\end{equation}
where the function $\lambda_0(r)$ is bounded as $r$ tends to zero. 
\end{lemma}

Proof.
The functions $\psi _{k}(z)$ analytic in $| z-a_{k}| < r_{k}$ are represented by their Taylor series
\begin{equation}
\label{eq:Tay}
\psi _{k}(z) = \sum_{l=0}^{\infty}\alpha_l^{(k)} \left(\frac{z-a_k}r\right)^l,
 \; | z-a_{k}| <r_{k}, \; k =1,2, \ldots,n.
\end{equation}
Here, the coefficients $\alpha_l^{(k)}$ are normalized in such a way that they are bounded as $r \to 0$. For definiteness, the eigenfunctions are supposed to be normalized as 
\begin{equation}
\label{eq:norm}
\|\phi\|_{\mathcal H^2}^2= \sum_{m=1}^n \sum_{l=0}^{\infty} \nu^l |\alpha_l^{(m)}|^2=1,
\end{equation}  
where $\phi(z)=\psi_{m}(z)$ in $| z-a_{m}|\leq r_{m}$.

Using \eqref{eq:Tay} we write equation \eqref{eq:rh14} up to $O\left(\frac{r^4}{\lambda(r)}\right)$  considering $\lambda(r)$ in general form since its asymptotic behavior has been not known yet
\begin{equation}
\begin{array}{lll}
\label{eq:rh92}
\psi_k(z)= r^2 \sum_{m\neq k}  \frac{\rho_m \nu_m}{(z-a_m)^2}  \left[
\overline{\alpha_0^{(m)}} +r\frac{\nu_m\overline{\alpha_1^{(m)}}}{z-a_m} \right] +
\\
\\
 \frac{r^2}{\lambda(r)} \sum_{m=1}^n \frac{\overline{\rho_m} \nu_m}{(1-\overline{a_m}z)^2}\; 
\left[\alpha_0^{(m)}+ r\frac{\nu_m \alpha_1^{(m)} z}{1-\overline{a_m}z} \right] +O\left(\frac{r^4}{\lambda(r)}\right),
\\
\quad \quad \quad \quad \quad \quad \quad \quad \quad \quad \quad \quad \quad \quad | z-a_{k}| \leq r_{k}, \; k=1,2,\ldots,n.
\end{array}
\end{equation}
%
Substitute $z=a_k$ into \eqref{eq:rh92} and reduce the order of approximation to $O\left(\frac{r^2}{\lambda(r)}\right)$ 
\begin{equation}
\begin{array}{lll}
\label{eq:rh93}
\alpha_0^{(k)}= r^2\sum_{m\neq k} \frac{\rho_m \nu_m}{(a_k-a_m)^2} 
\overline{\alpha_0^{(m)}}+
\frac{r^2}{\lambda(r)} \sum_{m=1}^n  \frac{\overline{\rho_m}\nu_m}{(1-\overline{a_m}a_k)^2}  \alpha_0^{(m)} +O\left(\frac{r^3}{\lambda(r)}\right),
\\
\quad \quad \quad \quad \quad \quad \quad \quad
\quad \quad \quad \quad \quad \quad \quad \quad \quad \quad \quad \quad \quad \quad  
k=1,2,\ldots,n.
\end{array}
\end{equation}
Differentiate equations \eqref{eq:rh92} and substitute $z=a_k$ into the result multiplied by $r$
\begin{equation}
\begin{array}{lll}
\label{eq:rh94}
\alpha_1^{(k)}= -2r^3\sum_{m\neq k} \frac{\rho_m \nu_m}{(a_k-a_m)^3}  \overline{\alpha_0^{(m)}}+
2\frac{r^3}{\lambda(r)} \sum_{m=1}^n  \frac{\overline{\rho_m}\nu_m  \overline{a_m} }{(1-\overline{a_m}a_k)^3} \alpha_0^{(m)}  +O\left(\frac{r^4}{\lambda(r)}\right),
\\
\quad \quad \quad \quad \quad \quad \quad \quad
\quad \quad \quad \quad \quad \quad \quad \quad \quad \quad \quad \quad \quad \quad  
k=1,2,\ldots,n.
\end{array}
\end{equation}
This procedure can be continued to get the next equations for $\alpha_l^{(k)}$ ($l=3,4,\ldots$).

We now prove that $\frac{r^2}{\lambda(r)}$ cannot tend to zero as $r\to 0$. If it is not so, then \eqref{eq:rh93} implies that $\alpha_0^{(k)}$ tends to zero as $r\to 0$. Then, equation \eqref{eq:rh94} implies that $\alpha_1^{(k)}$ tends to zero as $r \to 0$ and so forth $\alpha_l^{(k)} \to 0$ for all $l$. 
This contradicts to the normalization \eqref{eq:norm}.

The lemma is proved.\newline

It follows form Lemma \ref{lemma1} that the maximally possible absolute value of an eigenvalue for sufficiently small $r$ can be found in the form $\lambda = r^2 \mu +o(r^2)$, where $\mu$ is a non zero constant. Take the main terms of \eqref{eq:rh93} and write equations up to $O(r)$
\begin{equation}
\label{eq:rh101}
\mu \alpha_0^{(k,0)}= 
\sum_{m=1}^n  \frac{\overline{\rho_m}\nu_m}{(1-\overline{a_m}a_k)^2} \alpha_0^{(m,0)},
\quad  
k=1,2,\ldots,n,
\end{equation} 
where $\alpha_0^{(k,0)}=\alpha_0^{(k)}+O(r)$. Introduce the matrix $F$ whose elements have the form
\begin{equation}
\label{eq:rh102}
F_{mk}= \frac{\overline{\rho_m}\nu_m }{(1-\overline{a_m}a_k)^2}.
\end{equation} 
The eigenvalues $\mu$ of the linear algebraic system \eqref{eq:rh101} are solution of the polynomial equation 
\begin{equation}
\label{eq:rh103}
\det(\mu I- F)=0,
\end{equation} 
where $I$ stands for the identity matrix. 

If $\rho_m=\rho \in \mathbb R$ for any $m$, the matrix \eqref{eq:rh102} is self-adjoint. In this case, equation \eqref{eq:rh103} has exactly $n$ real roots counted with multiplicity.

The eigenfunctions can be constructed up to $O(r)$ by \eqref{eq:rh92}. Let $\mu$ be a simple eigenvalue and $\mathbf v =(\alpha_0^{(1,0)}, \alpha_0^{(2,0)}, \ldots, \alpha_0^{(n,0)})$ be the corresponding eigenvector of the linear algebraic system \eqref{eq:rh101}. Then, \eqref{eq:phi0}-\eqref{eq:phi01} yield
\begin{equation}
\label{eq:phi104}
\varphi_0(z) = -r^2\sum_{m=1}^n  \frac{\rho_m \nu_m}{z-a_m}\;\overline{\alpha_0^{(m,0)}},\quad |z| \geq 1.
\end{equation}

\begin{Example}[\cite{MitThesis}]
\label{example}
Let $n=1$ and $\Gamma_1 = \{t \in \mathbb C: |t|=r\}$ with $0<r<1$ in the problem \eqref{eq:pr1}-\eqref{eq:pr3}. All solutions of this problem have the following form 
 \begin {equation}
\label {prG3}
\varphi_1^{(p)}(z) =z^p, \; \varphi_0^{(p)}(z) =-\frac{1}{z^p}, \;  \varphi^{(p)}(z) =z^p-\frac{\rho r^{2p}}{z^p}, \; \lambda_p= \overline{\rho} r^{2p}, \; p=1,2, \ldots,\overline{\rho} 
\end {equation}
where the normalization $\|\varphi_1^{(k)}\|_{\mathcal H^2}^2=1$ is chosen in accordance with \eqref{eq:norm}.
\end{Example}

The case $p=1$ in Example \ref{example} corresponds to \eqref{eq:phi104} with $n=1$, $a_1=0$ and $\alpha_0^{(m,0)}=1$. The function, important in applications to metamaterials, $\omega_1(z) = \overline{\varphi_0^{(1)}\left(\frac{1}{\overline{z}}\right)}=z$ is univalent in the unit disk and determines a circle neutral inclusion with an annulus coating with a conductivity determined by $\lambda_1=\overline{\rho} r^2$ \cite{JM}.

\begin{Example}
\label{example2} 
Let $n=2$, $a_1=a$, $a_2=-a$ where $a$ be a positive number; $\Gamma_1 = \{t \in \mathbb C: |t-a|=r\}$ and $\Gamma_2 = \{t \in \mathbb C: |t+a|=r\}$ where $a+r<1$; $\rho_{1}=\rho_{2} \equiv \rho$. In this case, the system \eqref{eq:rh101} becomes
\begin{equation}
\label{eq:rh201}
\begin{array}{lll}
\mu \alpha_0^{(1,0)}= \overline{\rho} \left[\frac{1}{(1-a^2)^2} \alpha_0^{(1,0)} + \frac{1}{(1+a^2)^2} \alpha_0^{(2,0)} \right],
\\
\\
\mu \alpha_0^{(2,0)}= \overline{\rho} \left[\frac{1}{(1+a^2)^2} \alpha_0^{(1,0)} + \frac{1}{(1-a^2)^2} \alpha_0^{(2,0)} \right].
\end{array}
\end{equation}
The eigenvalues and eigenvectors of \eqref{eq:rh201}  have the form 
\begin{equation}
\label{eq:rh202}
\mu_1= \overline{\rho} \frac{2(1+a^4)}{(1-a^4)^2},\; \mathbf v_1=(1,1); \quad 
\mu_2= \overline{\rho} \frac{4a^2}{(1-a^4)^2},\; \mathbf v_2=(-1,1).
\end{equation}
The corresponding functions $\omega_p(z) = \overline{\varphi_0^{(p)}\left(\frac{1}{\overline{z}}\right)}$ are given by the approximate analytical formulae up to a multiplier $$\omega_1(z)=-\rho r^2\frac{2z}{1-a^2z^2}, \quad \omega_2(z)=\rho r^2 \frac{2az^2}{1-a^2z^2}.$$
One can see that the function $\omega_1(z)$ is univalent in the unit disks. It corresponds to the maximal $|\lambda_1|=r^2 |\mu_1|$.
\end{Example}

\section{Discussion}
The above study and examples enables us to make the following \newline
 
{\bf Conjecture}. {\it Let $\rho_k$ be given real numbers. Then, all eigenvalues of the problem \eqref{eq:pr1}-\eqref{eq:pr3} are real. The set of eigenvalues is countable or finite. Let $|\lambda_1| \geq |\lambda_2|  \geq \ldots $.  Then the corresponding eigenfunctions $\omega_p(z) = \overline{\varphi_0^{(p)}\left(\frac{1}{\overline{z}}\right)}$ ($p=1,2,\ldots$) satisfy inequality
 \begin {equation}
\label{prG2}
wind_{|z|=1} \omega_p(z) \leq p,
\end {equation}
where the winding number (or index \cite{Gakhov}) is defined as
$$
wind_{|z|=1} f(z) = \frac{1}{2 \pi i} \int_{|z|=1} \frac{f'(z)}{f(z)} \; dz.
$$}\newline

One can see in Example \ref{example} that 
$$wind_{|z|=1}  \omega_p(z)=wind_{|z|=1} \varphi^{(p)}_0(z)= p.$$ 
Moreover, $\max_k |\lambda_k|= |\lambda_1|$ and only the corresponding eigenfunction $\varphi_0^{(1)}(z)$ is conformal in $|z|>1$. 

Demonstration of Conjecture for $p=1$ allows to justify that any shaped inclusion with a smooth boundary can be made neutral by surrounding it with an appropriate coating \cite{JM}.

Conjecture recalls Courant's theorem \cite{Chavel} outlined below. Consider for definiteness the Dirichlet problem $u=0$ on $\partial \Omega$ for equation $\Delta u = - \lambda u$ valid in a domain $\Omega$. The set of eigenvalues consists of a sequence $0\leq \lambda_1 \leq \lambda_2 \leq \ldots $ (see Remark on page \pageref{rem1}) and the corresponding eigenfunctions $u_1, u_2, \ldots$ constitute a complete orthonormal basis of $L_2(\Omega)$. 
The nodal set of a fixed $u_p$ is defined as the set $\{z \in \Omega: \;u_p(z)=0\}$. According to Courant's theorem \cite{Chavel} the number of nodal domains of $u_p$ is less than or equal to $p$, for every $p=1,2,\ldots$. 

Conjecture can be stated in terms of nodal domains of the eigenfunctions $Re \; \varphi_0^{(p)}(z)$ in $|z|>1$ of the problem \eqref{eq:pr1}-\eqref{eq:pr3}. Instead of (\ref{prG2}) one can demand that the number of nodal domains of $Re \; \varphi_0^{(p)}(z)$ is less than or equal to $2p$, for every $p=1,2,\ldots$ . 

Let $\theta \in [0, 2 \pi)$ denote the argument of the complex number $z$.
It is easily seen that the nodal domains of the eigenfunctions $Re \; z^{-p} = |z|^{-p} \cos p\theta$ from Example \ref{example} are $2p$ sectors separated by the rays $\arg z = \frac{\pi m}p$ where $m=0,1, \ldots, 2p-1$. 

The general problem \eqref{R1}-\eqref{R3} and its partial case \eqref{eq:pr1}-\eqref{eq:pr3} for general curves $\Gamma_k$ have been not studied yet. Even in the case of $n$ sufficiently small circular inclusions Conjecture has been not proven.  It is reduced to the following seemingly simple question. Let points $a_k$ ($k=1,2,\ldots,n$) lie in the open unit disk and $\mathbf v=(\alpha_0^{(1,0)}, \alpha_0^{(2,0)},\ldots, \alpha_0^{(n,0)})$ be eigenvectors of the eigenvalue problem \eqref{eq:rh101}. For which $\mathbf v$ is the function 
\begin{equation}
\label{eq:phi0q}
\varphi_0(z) = \sum_{m=1}^n  \frac{\rho_m r^2_m}{z-a_m}\;\overline{\alpha_0^{(m,0)}}
\end{equation}
univalent in $|z|> 1$ or $\omega(z) = \overline{\varphi_0\left(\frac{1}{\overline{z}}\right)}$ in $|z| < 1$? Does this $\mathbf v$ correspond to the maximal $|\mu|$? This answer is interesting even for equal $\rho_m$ and $r_m$ when the number of eigenvectors holds $n$. It solves the problem of clouds of neutral inclusions. 

\section*{Acknowledgement}
The author thanks Ross C. McPhedran for stimulated discussions.


\begin{thebibliography}{99}

\bibitem{Alu}
A. Alu, and N. Engheta, Achieving transparency with plasmonic and metamaterial coatings, Physical Review E, 72, 016623, 2005. 

\bibitem{Asatryan}
Asatryan AA, Botten LC, Fang K, Fan S, McPhedran RC, Two-dimensional Green's tensor for gyrotropic clusters composed of circular cylinders, J Opt Soc Am A Opt Image Sci Vis. 2014  31, 2294-303. doi: 10.1364/JOSAA.31.002294.

\bibitem{Banuelos}
R. Banuelos, T. Kulczycki, I. Polterovich, B. Siudeja, Eigenvalue inequalities for mixed Steklov problems, arXiv:0909.5473, 2009.

\bibitem{Bogatyrev1}
Poincar\'{e}-Steklov integral equations and the Riemann monodromy problem, Functional Analysis and Its Applications, 2000, 34, 86-97.

\bibitem{Bogatyrev2}
A. Bogatyrev,
Poincar\'{e}-Steklov integral Equations and Moduli of Pants, Analysis and Mathematical Physics Trends in Mathematics 2009, 21-48

\bibitem{Bojarski}
B. Bojarski, On generalized Hilbert boundary-value problem, Soobsch. Akad. Nauk Gruz. SSR, 25, No. 4, 385-390 (1960).

\bibitem{BM}
B. Bojarski, V. Mityushev, R-linear problem for multiply connected domains and alternating method of Schwarz, Journal of Mathematical Sciences, 189, 68-77, 2013.

\bibitem{Chavel}
I. Chavel, {\it Eigenvalues in Riemannian Geometry},
Academic Press, London 1984.

\bibitem{McPh}
D J Colquitt, I S Jones, N V Movchan, A B Movchan, M Brun, R C McPhedran, Making waves round a structured cloak: lattices, negative refraction and fringes, Proc. Roy. Soc. A 2013; 469 20130218. DOI:10.1098/rspa.2013.0218

\bibitem{Drygas}
P. Drygas,
A functional-differential equation in a class of analytic functions and its application, Aequationes Math. 73 (2007) 222-232

\bibitem{Gakhov} 
F.D. Gakhov, {\it Boundary Problems}, Nauka, Moskow 1977 (in
Russian).


\bibitem{JM}
P. Jarczyk, V. Mityushev, Neutral coated inclusions of finite conductivity, {\it Proc. R. Soc. Lond.}, {\bf A} (2011), 2011 Published online doi:10.1098/rspa.2011.0230.

\bibitem{andr2}
 A.L. Kalamkarov,  I.V. Andrianov,  V.V. Danishevskyy, Asymptotic homogenization of composite materials and structures, {\it Appl. Mech. Rev.}, {\bf 62} (2009), no. 3, 030802-1--030802-20 

\bibitem{Kerker}
 M. Kerker, Invisible bodies, {\it J. Opt. Soc. Am.} {\bf 65} 1975, 376--379.

\bibitem{KF}
A.N. Kolmogorov and S.V. Fomin, Elements of the Theory of Functions and Functional Analysis, Dover Publications, NY, 1999  
 
\bibitem{ag}
V. I. Lebedev, V. I. Agoshkov, {\it Poincar\'{e} - Steklov operators and their applications in analysis}, Akad. Nauk SSSR, Vychisl. Tsentr, Moscow 1983 (in Russian).


\bibitem{MiltonS}  G. W. Milton,  S. K. Serkov, Neutral coated inclusions in conductivity and anti-plane elasticity, {\it Proc. R. Soc. Lond.}, {\bf  457} ( 2001), 1973--1997.

\bibitem{MitThesis} 
V. Mityushev, Boundary value problems and functional equations with shifts in domains, PhD Thesis, Minsk 1984 (in Russian).

\bibitem{Mityushev}  V.  Mityushev, Eigenvalues of the $\mathbb{R}$--linear problems, {\it Izvestia vuzov. Math.,} {\bf 11} (1992),  35-38 (in Russian). 

\bibitem{Mityushev3} V.  Mityushev, S. Rogosin, {\it Constructive Methods for Linear and Nonlinear Boundary Value Problems for Analytic Functions. Theory and Applications}, Chapman \& Hall / CRC, Boca Raton etc 2000.

\bibitem{3}
V.V. Mityushev, Riemann-Hilbert problems for multiply connected domains and circular slit maps, Comp. Meth. Func. Theory, 11 (2011), 557-590.


\bibitem{Mit2012b}
V. Mityushev, N. Rylko, Optimal distribution of the non-overlapping conducting disks. Multiscale Model. Simul., 10  (2012), 180-190.


\bibitem{McPhedran2015}
J. O'Neill, \"{O}. Selsil, R.C. McPhedran, A.B. Movchan, and N.V. Movchan,
Active cloaking of inclusions for flexural waves in thin elastic plates
Q J Mechanics Appl Math first published online June 25, 2015
doi:10.1093/qjmam/hbv007.

\bibitem{Poulton}
C.G. Poulton, A.B. Movchan, N.V. Movchan, R.C. McPhedran, 
Analytic theory of defects in periodically structured elastic plates, Proc. R. Soc. A471, 2012, 1196-1216. DOI: 10.1098/rspa.2011.0609

\bibitem{Rylko0}
N. Rylko, Transport properties of a rectangular array of
highly conducting cylinders, J. Engineering Math. 38, (2000) 1-12.

\bibitem{Rylko1}
N. Rylko, Structure of the scalar field around
unidirectional circular cylinders, Proc. R. Soc. A464 (2008), 391–407.

\bibitem{Rylko2}
N. Rylko, Effective anti-plane properties of piezoelectric fibrous
composites, Acta Mech 224, 2719–2734 (2013)

\bibitem{Rylko3}
N. Rylko, Fractal local fields in random composites, Computers and Mathematics with Applications 69 (2015) 247-254.

\bibitem{Sch}  M. Schiffer, Fredholm eigen values of multiply-connected domains, {\it Journal d'Analyse Math\'{e}matique}, {\bf 9} (1959), 211--269.

\end{thebibliography}
\end{document}